# Generalized 3D Voxel Image Synthesis Architecture for Volumetric Spatial Visualization

**Anas M. Al-Oraiqat *[1], E. A. Bashkov [2], S. A. Zori [2], Aladdein M. Amro [3]**

*[1] Taibah University, Department of Computer Sciences and Information
Kingdom of Saudi Arabia, P.O. Box 2898

[2] SHEE «Donetsk National Technical University», Ukraine, P.O.Box 85300

[3] Taibah University, Department of Computer Engineering,
Kingdom of Saudi Arabia

**ABSTRACT**

*A general concept of 3D volumetric visualization systems is described based on 3D discrete voxel scenes (worlds) representation. Definitions of 3D discrete voxel scene (world) basic elements and main steps of the image synthesis algorithm are formulated. An algorithm for solving the problem of the voxelized world 3D image synthesis, intended for the systems of volumetric spatial visualization, is proposed. A computer-based architecture for 3D volumetric visualization of 3D discrete voxel world is presented. On the basis of the proposed overall concept of discrete voxel representation, the proposed architecture successfully adapts the ray tracing technique for the synthesis of 3D volumetric images. Since it is algorithmically simple and effectively supports parallelism, it can efficiently be implemented.*

## 1.INTRODUCTION

The modern virtual reality systems (VRS) very often use various methods of spatial 3D visualization for spatial representation of data. One of the main applications of the VRS is the production of training devices and simulators especially for training in potentially dangerous areas of human activities [1].

The environment visualization systems (EVS) of high quality, expensive, industrially produced simulators are mainly being constructed todays, making use of projection and collimation screens [1]. It is impossible to conduct geometrically correct visual simulation due to the considerable shortcomings of the available flat and spherical image projection devices and image generation systems, such as parallax errors and monocular principle of 3D image synthesis, as well as the high cost of modern collimation spherical screens and EVS. Accordingly, intended to increase the degree of the synthesized and visualized image realism, 3D volumetric synthesis and visualization are considered to be a perspective approach.

The analysis of volumetric spatial visualization methods and means has made it possible to define their main types and development tendencies and to formulate the main requirements of visualization systems, based on their principles [1-4]. The increasing practicality of graphic scenes spatial 3D synthesis, in virtual modeling systems and various types of simulators, principally provides new process organization and 3D pipelining, making use of complex imaging methods and synthesis such as ray tracing. It also develops modifications for spatial volumetric synthesis [1,5].

Nowadays two basic ways for organizing the spatial 3D visualization systems are mainly used: stereoscopic 3D visualization and volumetric spatial 3D visualization [1]. The majority of 3D information representation systems (3D IRS) and 3D displays make use of stereoscopic 3D visualization because of its easier and cheaper realization. The systems of the given class represent "reasonable compromise" between the quality, speed and price of 3D visualization. However, to achieve a desirable degree of photorealistic image quality in such systems, it is necessary to use ray tracing methods, which are extremely labor consuming and do not directly use procedures of standard 3D graphic pipeline and standard processor software of graphic sub-systems [1-5]. Therefore, raising the speed and synthesis quality in such systems is a very practically relevant and promising task [1, 2, 4].

So far the systems of volumetric spatial 3D visualization and displays don't permit to obtain full or high quality 3D images, as their prices are high and they still have a limited circle of users. The most popular among them are the systems of voxel representation and visualization.

These systems, however, are nowadays characterized by absence of 3D information representation standardization since the definitions of typical 3D graphic primitives/algorithms for their generation have not been worked out yet. So the development of volumetric approximation and 3D object generation methods are essentially of great importance [1, 2, 5]. Additionally, the operation of dynamic scene synthesis in real time still requires more effective production of 3D graphics IRS [1, 4]. Consequently, the rendering task of 3D volumetric images in the systems of volumetric 3D spatial visualization, based on voxel representation of the scenes and ray tracing methods, is very crucial.

## 2. 3D DISCRETE WORLD REPRESENTATION AND IMAGE SYNTHESIS

The systems of volumetric spatial visualization, based on voxel representation and 3D volumetric scene visualization, has recently been considered to be very promising but, at the same time, insufficiently investigated [1,5-8]. The following generalized concept of discrete volumetric 3D world representation and 3D volumetric image synthesis is proposed as illustrated in Figure 1.

First, assume the following:
1) **Voxel** – a minimum 3D spatial element of some 3D space, discretized at a definite pitch. In general case, it is a 3D rectangular parallelepiped. In the simplest case, it is a 3D unity cube with unitary sides.
2) **Voxelized discrete world/scene** – volumetric voxel representations (models, approximations) of the objects and their surrounding 3D world space:

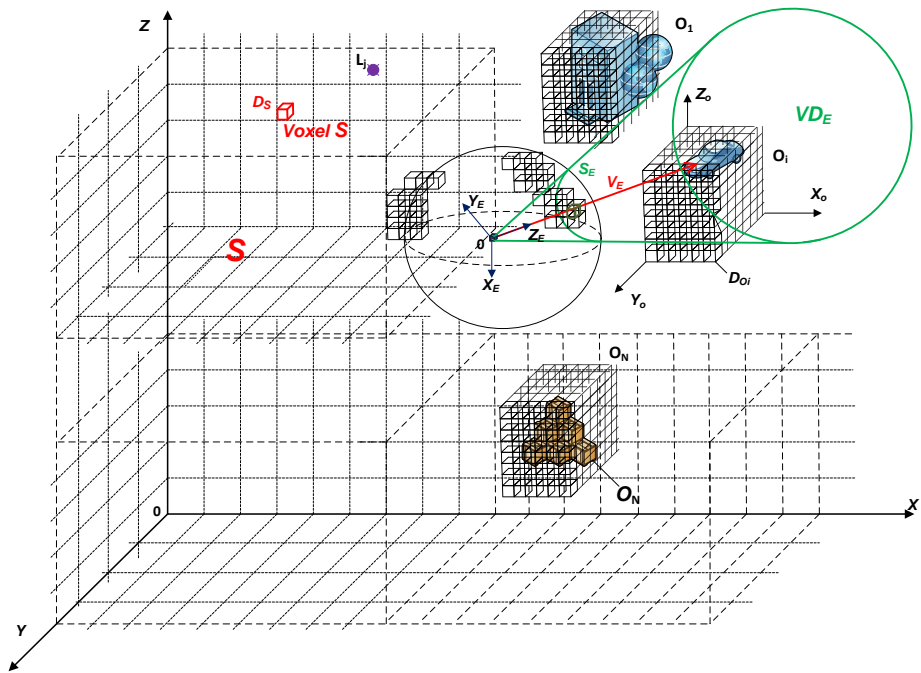

**Figure 1** Concept of discrete 3D volumetric world and image synthesis in 3D volumetric IRS.

2.1) **Parameters of the world spatial discretization** – scene space resolution $Ds$ and object space resolution $Do$.
2.2) **World coordinate system (WCS) – $OXYZ$**. In WCS voxel scene space representation $S$ is defined as $\mathbb{S}$ array of $S_{l,m,n}$ voxels, where $l$, $m$ and $n$ are voxel indices in $\mathbb{S}$, corresponding to coordinates $XYZ$ in WCS.
The value of $S_{l,m,n}$ voxel sets a certain "space" characteristics of the scene space in the volume, defined by the given voxel (e.g. absorption ratio, etc.)
2.3) **Voxel representation of the scene objects $O_i$, defined within the object coordinate system (OCS) $OXo_iYo_iZo_i$** or within WCS.
$O_i$ object representation is a multitude array $\mathbb{O}_i$ of $\mathcal{O}_{l,m,n}$ voxels, so that $\mathcal{O}_{l,m,n} \in O_i$, and $l$, $m$ and $n$ are voxel indices in $\mathbb{O}_i$, corresponding to $Xo_iYo_iZo_i$ coordinates in OCS [8]

$O_i$ is given its own (associated) coordinate system ($O_i$CS):
- ($X_0o_i, Y_0o_i, Z_0o_i$) depending on position of $O_i$CS in relation to WCS.
- ($\psi_{oi}, \theta_{oi}, \gamma_{oi}$) depending on turn angles of $O_i$CS in relation to WCS.

In this case the value of $\mathcal{O}_{l,m,n}$ voxel is a certain object characteristic within the given "elementary" volume (color, transparence, etc.).

2.4) **Sources of illumination $L_j$** are set within WCS by means of their positions ($X_{Lj}, Y_{Lj}, Z_{Lj}$) and energy characteristics (color, intensity etc.).

It should be noted that the sources of light can be in the form of multiple spots. In this case, a geometric description of the source surface and its energy characteristics is considered for each source belonging to $O_i$.

3) **Observer – E**:

3.1) **Observer coordinate system ($O_v$CS) – $OX_EY_EZ_E$.**

$O_v$CS is given in relation to WCS like OCS – by means of its position ($X_{0E}, Y_{0E}, Z_{0E}$) and orientation ($\psi_E, \theta_E, \gamma_E$) as turn angles of the observer coordinate system in relation to the world coordinate system.

3.2) **Direction of the observer's view (sight ray) – $V_E$.** Usually the direction of visualization coincides with $Z_E$ axis.

3.3) **Sector of the observer's spatial observation – $S_E$.**

The observation sector can be affected by different means – by spatial angular characteristics, limiting rays and so on, depending on the used observer's model and form of display.

4) **Spatial screen (display)**

**Visible volume sight space – $VS_E$:** $VS_E$ is a spatial geometric part of the volumetric space $S$ of the world (scene), seen by the observer $E$, defined by the observation physical model, $VS_E \in S$. The visible volume $VS_E$ is set as an array $\mathbb{V}$ of $\mathcal{V}_{l,m,n}$, voxels and it fully fills (covers) $VS_E$.

**Visible volume display - $VD_E$:** $VD_E$ is a part of the volumetric world (scene) space – $VS_E$, seen by the observer $E$ (the display depicts the visible world space).

**Physical observation model:** Depending on a physical observation model (either making use of human eye sight specificity or not) or other accepted simplifications and limitations, one can define various forms of spatial volumetric displays as shown in Figure 2.

Possible forms of the sight space $VS_E$ of the volumetric display $VD_E$ are presented that take into account the human eye specificity (eye and sight model as a reverse projection) in Figure 2 (a) and (b), and the simplified models (without eye model and simplified sight space form) in Figure 2 (c) and (d). Depending on the size of the observation sector, visible volume and simplifications in observation characteristics, one can use spherical, cubic volumetric displays, volumetric displays with sight space in the form of truncated cones, rectangular pyramids, etc. [1].

Making use of volumetric special display idea, it is possible to define a volumetric image.

**Volumetric image**: visible for the observer volumetric part of voxelized discrete world, caught in the sight space of volumetric display as illustrated in Figure 3.

5) **World dynamic characteristics**:

5.1) **Parameters characterizing temporal behavior of the objects:**
- Changes in spatial discretized geometry representation of each scene object, namely $\mathbb{O}_i$ change.
- Changes of the objects visual characteristics (changes of $\mathbb{O}_i$ values).
- Changes of the objects disposition and orientation (movements, rotations, etc.) in WCS.

5.2) **Parameters of dynamics illumination sources:**
- Parameters of spatial disposition of illumination sources.
- Parameters of changes in illumination sources energy characteristics.

5.3) **Parameters of observer's dynamics:**
- Parameters of observer's spatial disposition and orientation.
- Parameters of changes in direction of viewing and observation sector.

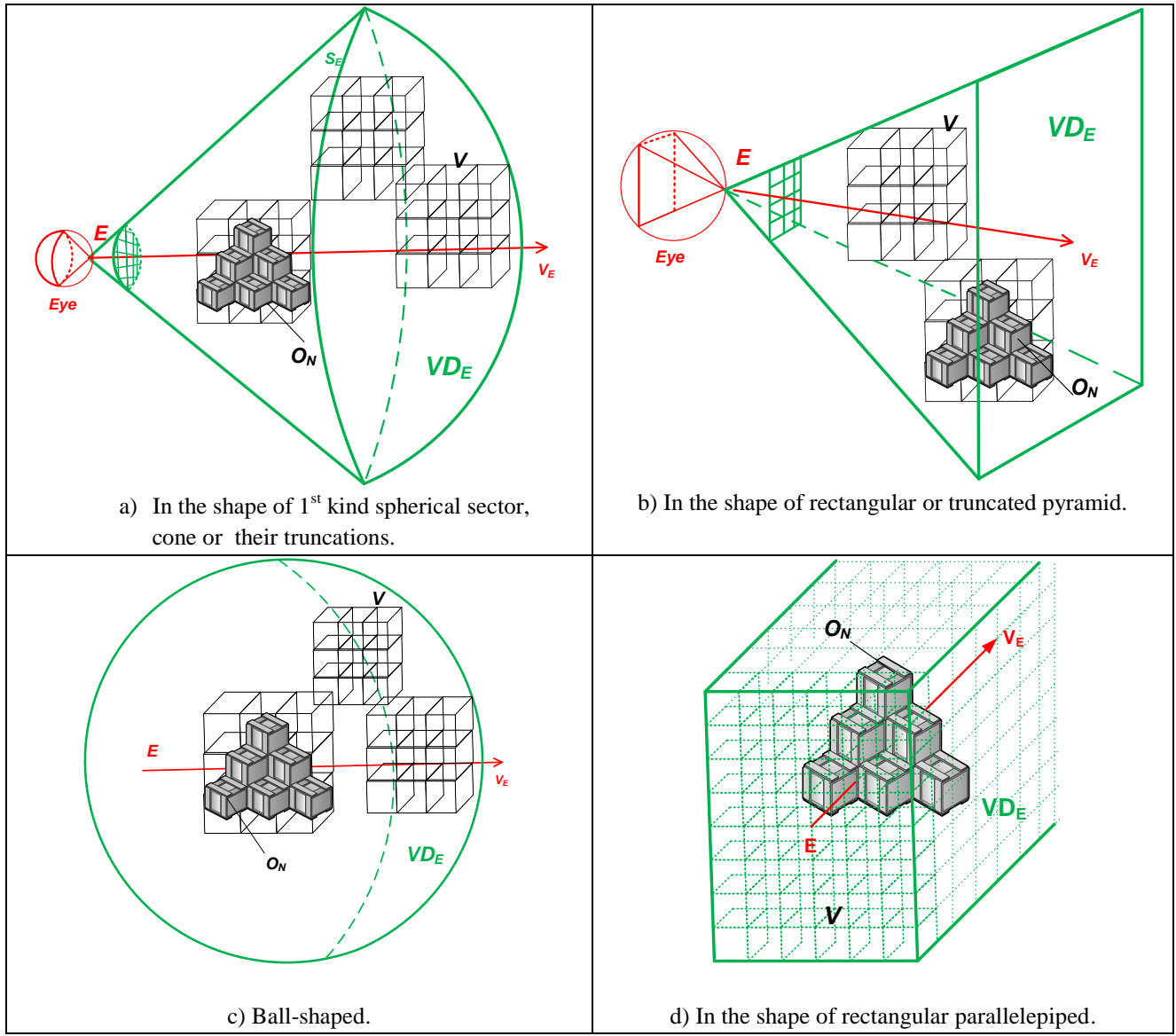

**Figure 2** Shapes of view space volumetric displays. (a) In the shape of 1st kind spherical sector, cone or their truncations. (b) In the shape of rectangular or truncated pyramid. (c) Ball-shaped. (d) In the shape of rectangular parallelepiped.

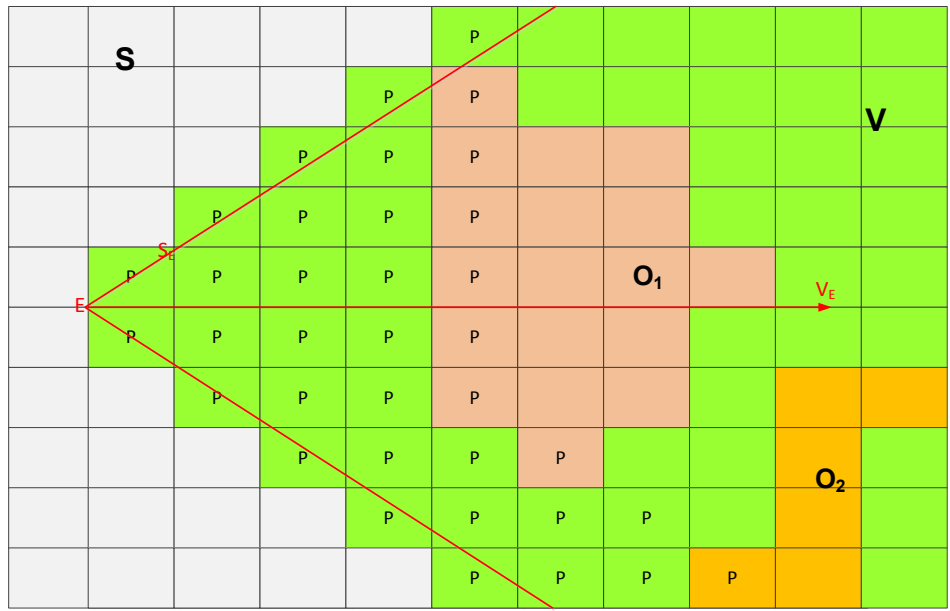

**Figure 3** Volumetric representation

5.4) **Dynamic parameters change of surrounding inner scene space**: Value changes of $S_{l,m,n}$.

It is a task of the volumetric visualization to build up a volumetric spatial-discretized world image, volumetric representation $\mathbb{P}$, seen by the observer from his/her position in the direction of viewing in the space of volumetric display. It means to find such a voxel multitude $\mathfrak{P}_{l,m,n}$ for each moment of time that $P_{l,m,n} \in P$, $\mathbb{P} = \mathbb{V} \cap \mathbb{S} \cap_{i=1}^{N} \mathbb{O}_i$ as shown in Figure 3. Each voxel of the volumetric representation $\mathfrak{P}_{l,m,n}$ has to obtain a characteristic value $C_{\mathfrak{P}_{l,m,n}}$, taking into account voxel characteristics $C_{\mathcal{O}_{l,m,n}}$ of the objects $C_{\mathcal{O}_{l,m,n}}$, falling into specific volume $VS_E$ of the display of $C_{L_j}$ illumination sources and of voxel scene space $C_{S_{l,m,n}}$

$$C_{\mathfrak{P}_{l,m,n}} = C_{\mathcal{O}_{l,m,n}} \cup\ C_{L_j}\ \cup C_{S_{l,m,n}}.$$

It is worth to mention that the main particularity of such an approach to volumetric spatial visualization is as follows. The image construction can be brought to the definition of the object voxels model, visible to the observer, within a specific spatial volume volumetric display. Moreover, the illumination parameters and other visual properties of the inner surrounding world space are considered.

### 3. SYNTHESIS ALGORITHM OF VOXEL WORLD VOLUMETRIC REPRESENTATION

To synthesize a volumetric representation of the voxelized world within the framework of the proposed concept, with the preset models of the objects in WCS, it is sufficient to do the following:

1. Construct a world voxel model: A voxel model represents the surrounding space and objects with their spatial disposition in the scene. If the geometry and the object optical properties don't change, the calculations at this stage are conducted as preprocessing once.
2. Conduct sight ray tracing procedure at the given step of spatial discretizing in the scene volume space, visible to the observer, until the first intersection with the voxel of some object, directly seen by the observer in the given point.
3. Define visual characteristics of the found voxel. It is necessary to carry out ray tracing to the sources of light and to correctly "mix up" the voxel proper color and its optical characteristics with the characteristics of the visible sources of light and the environment.
4. Possibly compute more complex illumination characteristics for the found voxel object element, such as photorealistic effects of refractions and reflections, making use of analogous, used in traditional 3D graphics ray tracing methods [1,3,5-7,9,10].

The proposed generalized algorithm of discrete voxel world volumetric representation is as follows:

**Begin**
  Form $\mathbb{S} = \{S_{l,m,n}; C_{S_{l,m,n}}\}$ ;
    **Do** (for all $O_i$ objects of the volumetric world $S$):
      Form $\mathbb{O}_i = \{O_{l,m,n}; C_{O_{l,m,n}}\}$ ;
      Transform $O_i$ in WCS and place it in $S$: $\mathbb{S} \cup \mathbb{O}_i$ ;    → **1**
    **End_Do**
  **Do** (for all $\mathbb{V}$ voxels of volumetric screen $VD_E$):
    $\mathcal{V}_{l,m,n} = S_{l,m,n}; C_{\mathcal{V}_{l,m,n}} = C_{S_{l,m,n}}$ ;
  **End_Do**
  **Do** (for each sight ray of specific space $VS_E$ (of the initial ray) from $E$ with given $D_S$ and $S_E$):
    Conduct discrete reflected ray tracing for each step ($k$) via $VS_E$ voxels;   → **2**
    Find the first intersection of the ray with voxel of some world object $O_i$ ;   → **3**
    $\mathfrak{P}_{l,m,n}(k) = \mathcal{V}_{l,m,n}(k); C_{\mathfrak{P}_{l,m,n}}(k) = C_{\mathcal{V}_{l,m,n}}(k)$ ;
    **If** (there is intersection) **Then**
      $\mathfrak{P}_{l,m,n}(k) = \mathcal{O}_{l,m,n}(k)$ ;
      **Do** (for each light source $L_j$):
        Find intersection of the initial ray with the object;   → **4.1**
        Conduct discrete shade ray tracing for each step from the voxel of the initial ray
        intersection with the object towards the light source $L_j$ ;   → **4.2**
        **If** (there is intersection)
          **If** (intersection with some world object $\mathbb{O}_i'$) **Then**
            Stop processing this $L_j$ (it is not seen), voxel is in shade;
          **Else**
            Take into account the degree of this source illumination;   → **5**
            $L_j: C_{\mathfrak{P}_{l,m,n}}^{\ j} = C_{\mathcal{O}_{l,m,n}}(k) \cup C_{L_j}$;
          **End_If**
        **End_If**
        $C_{\mathfrak{P}_{l,m,n}}(k) = C_{\mathcal{V}_{l,m,n}}(k) + C_{\mathfrak{P}_{l,m,n}}^{\ j}$;
      **End_Do**
      **If** ($\mathbb{O}_i$ object voxel material is reflecting) **Then** Conduct refracted ray tracing;   → **6.1**
        Find the first intersection of the ray with the voxel of some world object $\mathbb{O}_i''$;   → **6.2**
        **If** (there is intersection with some object $\mathbb{O}_i''$)
          Take into account $\mathbb{O}_i''$ object voxel characteristic with ratio $R_i$ of the material $\mathbb{O}_i$;   → **7**
          $C_{\mathfrak{P}_{l,m,n}}(k) = C_{\mathfrak{P}_{l,m,n}}(k) \cup (C_{O''_{l,m,n}} * R_i)$;
        **End_If**
      **End_If**
      **If** ($\mathbb{O}_i$ voxel material is refracting) **Then**
        Conduct refracted ray tracing;   → **8.1**
        Find the first ray intersection with the voxel of some $\mathbb{O}_i'''$ object of the world;   → **8.2**
        **If** (there is intersection with some $\mathbb{O}_i'''$ object)
          Take into account the value of $\mathbb{O}_i'''$ voxel object with refraction ratio $F_i$ of material $\mathbb{O}_i$;   → **9**
          $C_{\mathfrak{P}_{l,m,n}}(k) = C_{\mathfrak{P}_{l,m,n}}(k) \cup (C_{O'''_{l,m,n}} * F_i)$ ;
        **End_If**
      **End_If**
    **End_If**
  **End_Do**
**End**

The above proposed algorithm is a generalized algorithm of the problem solution, involving synthesis of volumetric representation of voxel discrete world in a volumetric 3D IPS. The algorithm includes simplified definitions of refractions and reflections (solitary effects, analogous to the base ray tracing algorithm of [9]). So, the proposed approach of combining spatial-discrete world representation and spatial discretized ray tracing makes it possible to construct the spatial representation in a quite simple and effective algorithmic way. It simultaneously solves a number of problems including stages of 3D graphic pipeline, scenario processing of geometric model changes, rendering with solutions of invisible surface removal and illumination problems, and proper visualization.

## 4. ARCHITECTURE OF DISCRETE VOXEL WORLD 3D VOLUMETRIC REPRESENTATION SYNTHESIS SYSTEM

Using the suggested concept of the discrete 3D volumetric world representation, the architecture of the proposed volumetric image synthesis system is shown in Figure 4. The underlying processes and/or procedures of each processing block are as explicitly described in the corresponding steps of the proposed algorithm presented in Section 3, with the associated numbers for their referencing. To accelerate the process of the dynamic volumetric world representation synthesis, it is possible to make use of known computer graphics approaches such as paralleling, temporal and spatial decomposition of the treatment processing, temporal and spatial decomposition of the processing.

The main features of the proposed architecture are as follows:
- It introduces the synthesis of new 3D voxel volumetric images.
- It allows the use of the existing technique of ray tracing on the basis of the proposed overall concept of discrete voxel representation of the world. It is also considered as an adaptation of the traditional architecture of ray tracing for the synthesis of 3D volumetric images that algorithmically constructs volumetric images using ray tracing.
- It effectively supports parallelism since all involved ray tracer problems can be solved independently.
- It is algorithmically simple enough to implement.

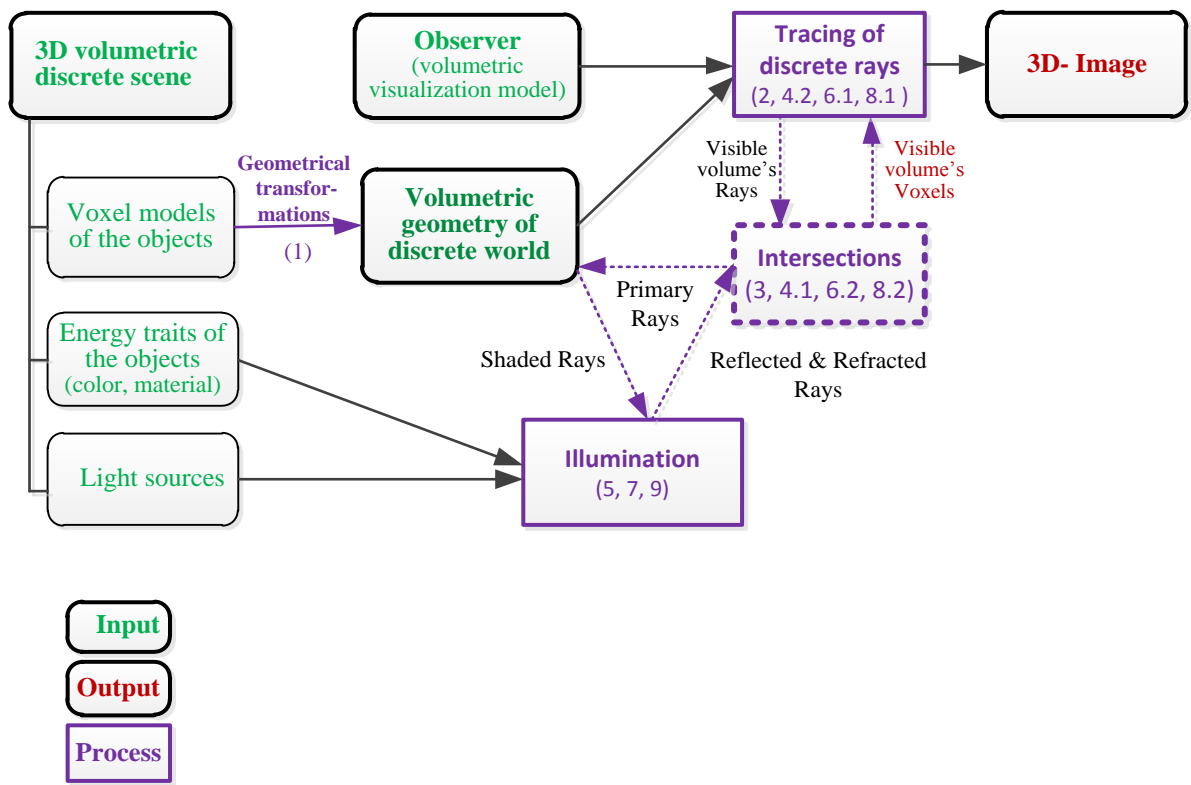

**Figure 4** Generalized architecture for synthesis of discrete voxel world 3D images.

For example, the volumetric display organization can be used with the frame buffers to maintain the preset volumetric representation synthesis criteria. Also the pipeline treatment can be used to shorten the time of the 3D spatial visualization. Alternatively, it is possible to attach the computing components of 3D IPS to the corresponding world objects in order to provide parallel restructuring of the object spatial voxel models for the geometric affinitive transformations.

To effectively realize the suggested concept of discrete voxel world volumetric representation synthesis, one must be able to:
- Productively generate voxel representations (models) of the spatial-discrete world objects.
- Carry out fast discrete ray tracing.

## 5. CONCLUSION

The article moves forward the concept of 3D volumetric world representations. A solution algorithm for the problem of the voxelized world volumetric representation synthesis, applied to the systems of the volumetric spatial visualization, has been suggested. The main steps of the image synthesis algorithm are formulated. A generalized architecture for synthesis of 3D volumetric discrete voxel world representations has been introduced. The basic features and advantages of the proposed dynamic volumetric image synthesis architecture are presented. On the basis of the proposed overall concept of discrete voxel representation of the world, the proposed architecture adapts the traditional technique of ray tracing for the synthesis of 3D volumetric images. Also, it can efficiently be implemented because it is algorithmically simple and effectively supports parallelism.

## ACKNOWLEDGMENT

The authors would like to thank Taibah University and Donetsk National Technical University for supporting this research.

## References

[1]. E. A. Bashkov and S. A. Zori, "Realistic spatial visualization using volumetric display technologies," Donetsk, DonNTU, p. 151, 2014.

[2]. E. A. Bashkov and S. A. Zori, "Application of graphic system computation capabilities for three dimensional scene visualization making use of volumetric representation technology," Izvestia UFU, Technical Sciences, 4 (153), pp. 15-21, 2014.

[3]. S. A. Zori and P. A. Porfirov, "Productivity increasing of realistic ray tracing stereo- image synthesis," Journal of Qafqaz University, Mathematics and Computer Science, Vol. 3, No.1, pp. 30- 38, 2015.

[4]. T. Akenine-Moller, E. Haines and N. Hoffman, "Real-time rendering," Wellesley: A. K. Peters, Ltd., 3$^{rd}$ Edition, p. 1027, 2008.

[5]. C. Crassin, F. Neyret, S. Lefebvre and E. Eisemann, "Ray-guided streaming for efficient and detailed voxel rendering" ACM SIGGRAPH Symposium on Interactive 3D Graphics and Games (I3D), Feb. 2009.

[6]. Arslanov Dmitriy Merzagitovich. "Method of voxel rusterization and processing," http://rsdn.org/article/alg/03-12-voxel.xml, (Accessed on 12 Oct., 2016).

[7]. S. Forstmann and J. Ohya, "Efficient, high-quality, GPU-based visualization of voxelized surface data with fine and complicated structures", IEICE Transactions on Information and Systems, Vol. E93-D, No.11, pp. 3088-3099, Nov. 2010.

[8]. E. A. Bashkov and S. A. Zori, "On voxel presentation of arbitrary spatial curve," International Conference on Modeling and Computer Graphics, Krasnoarmeisk, Ukraine, pp. 101-112, 25-29 May 2015.

[9]. T. Whitted, "An improved illumination model for shaded display," ACM Communications, Vol. 23 (6), pp. 343–349. 1980.

[10]. P. Shirley and R. K. Morley, "Realistic Ray Tracing," Wellesley: A K Peters/CRC Press, , 2$^{nd}$ Edition, p. 235, 2003.

## Author's Profile

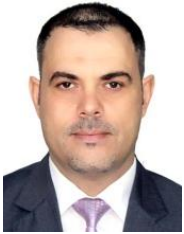
**Anas M. Al-Oraiqat** received a B.S. in Computer Engineering and M.S. in Computer Systems & Networks from Kirovograd Technology University in 2003 and 2004, respectively, and Ph.D. in Computer Systems & Components from Donetsk National Technical University (Ukraine) in 2011. He has been an Assistant Professor at the Computer & Information Sciences Dept., Taibah University (Kingdom of Saudi Arabia) since Aug. 2012. Prior to his academic career, he was a Network Manager at the Arab Bank (Jordan), 2011-2012. Also, he was a Computer Networks Trainer at Khwarizmi College (Jordan), 2005-2007.
His research interest is in the areas of computer graphics, image/video processing, 3D devices, modelling and simulation of dynamic systems, and simulation of parallel systems.
E-mail: anas_oraiqat@hotmail.com

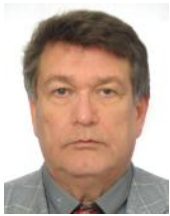
**Evgeniy A. Bashkov** received a Associate Professor in Computer Systems & Components from Pukhov Institute for Modeling in Energy Engineering National Academy of Science of Ukraine in 1995 and Professor in 1996. He is Professor of Applied Mathematics & Computer Science Department of the Faculty of Computer Science and Technologies of Donetsk National Technical University (DonNTU).
His research interest is in the areas of the real time computer graphics, virtual reality, the computer image generators, the simulation system, the high performance computing systems of common and special purpose.
E-mail: eab23may@gmail.com

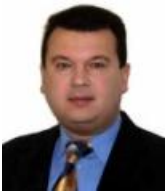
**Sergii A. Zori** received a Ph. D. in Computer Systems & Components from NTU "Donetsk National Technical University" (Ukraine) in 1996 and Associated Professor in 2017. He is Associate Professor at the Mathematics & Computer Science Department of the Faculty of Computer Science and Technologies, Donetsk National Technical University ("DonNTU").
His research interest is in the areas of computer graphics, image/video processing & vision, 3D devices, Virtual reality systems, specialized parallel computer systems.
E-mail: sa.zori1968@gmail.com

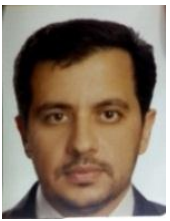
**Aladdein M. Amro** received M.S. in Automation Engineering from Moscow Technical University in1996, and Ph.D. in Telecommunications Engineering from Kazan State University (Russian Federation) in 2003. Had been an Assistant Professor at the Computer Engineering Dept., Al-Hussein Bin Talal University (Jordan) during the years 2004-2011. Since then has been working as an Assistant Professor at the Computer Engineering Dept., Taibah University (Kingdom of Saudi Arabia). Research interest is in the areas of digital Signal processing, image processing, real time systems.
E-mail: amroru@hotmail.com